# Do the short helices exist in the nematic TB phase?

Ewa Gorecka,* Miroslaw Salamonczyk, Anna Zep, Damian Pociecha

*Department of Chemistry, University of Warsaw, Zwirki i Wigury 101, 02-089 Warsaw, Poland.*
*e-mail: gorecka@chem.uw.edu.pl*

Chris Welch, Ziauddin Ahmed, Georg H. Mehl

*Department of Chemistry, University of Hull, Hull HU6 7RX, UK*

# Do the short helices exist in the nematic TB phase?

Dimeric compounds forming twist-bend nematic, $N_{tb}$, phase show unusual optical texture related to the formation of arrays of focal conic defects. Some of the focal conics show submicron internal structure with 8 nm periodicity, which is very close to that found in the crystalline phase of the material, suggesting surface freezing.

The nematic phase was considered for a long time as an educative example of the 'simplest' liquid crystalline phase, the phase is built of molecules having long range orientational order but lacking long range positional order. Though examples for the nematic-nematic phase transition have been shown for polymers [1] this simple picture was recently challenged significantly. It was shown that for some dimeric or bent-core molecules, more than one nematic phase exists [2-6], upon lowering temperature the 'classical' nematic phase with uniform orientation of the director transforms by a first order transition to a nematic phase with a spontaneous spatial modulation of the director. Chiral domain formation, associated with a very fast electro-optic response was also found [7-9] in the new nematic phase. Based on transmission electron microscopy (TEM) studies performed for replicas of freeze fractured samples [3,4], nuclear magnetic resonance (NMR) studies [10-11] and electro-optical investigations [9] a picture for the low temperature nematic phase was proposed as having short oblique helicoidal twist-bent (TB) modulations, with an extremely short helical pitch of the size of a few molecular lengths (8- 10 nm), in line with theoretical models [12-14]. However, the phase structure assignment is problematic, as such a short, regular structures visible in TEM images are not detectable by other direct methods. Moreover, recent results of NMR studies suggest that the $N_{tb}$ phase might not be composed of short twist–bend helices [15]. Here we show that periodic, submicron features can be ascribed to crystallographic planes of a solid crystal, easily formed during 'freezing' of the samples, so the alternative models for the structure of the $N_{tb}$ phase need to be explored.

Materials of the homologous series **CB-n-CB** (with two 4-cyanobiphenyl mesogenic cores linked by flexible alkyl spacer with n carbon atoms), showing the N-Ntb phase transition, studied previously by the Boulder [3] and Kent [4] groups, were re-investigated, mainly by atomic force microscopy (AFM) techniques. Samples were prepared in a similar manner as reported for transmission electron microscopy (TEM) studies [3,4], i.e. the material was placed between solid substrates (glass or metal), slowly cooled from the isotropic phase to the $N_{tb}$ phase, and then quickly immersed in liquid nitrogen. Subsequently, one substrate was removed and the free surface of material was studied by AFM at room temperature.

For the cyanobiphenyl dimers with an odd number of carbon atoms in the linking group (homologues with n = 7, 9 and 11), the N-$N_{tb}$ phase transition was easily detected by differential scanning calorimetry (DSC) and polarizing optical microscopy (POM). It was confirmed by XRD that both nematic phases have only short range positional order, with correlation length up to 1-2 molecular distances, in line with previous results [2]. The optical texture of the higher temperature phase is typical for the nematic phase; when placed between glass plates with uni-directionally rubbed aligning polymer layers, molecules are homogeneously oriented along the rubbing direction, the birefringence (close to the N-$N_{tb}$ transition) is $\Delta n=0.15$. The texture of the lower temperature phase ($N_{tb}$) shows a number of densely packed stripe-like defects aligned along the rubbing direction (Fig. 1).

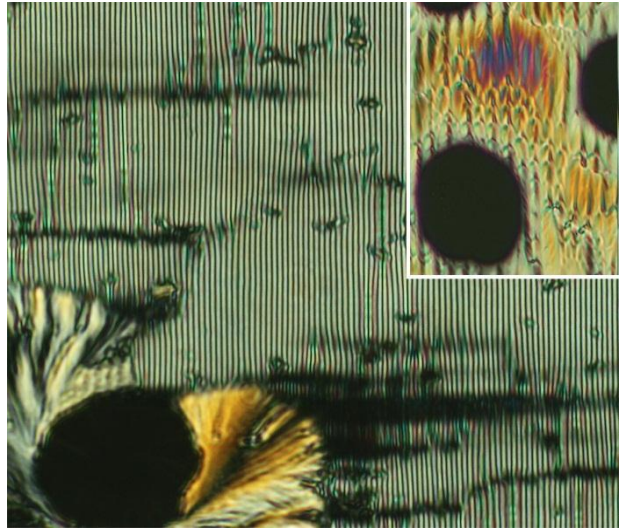

**Figure 1.** Stripe texture of the $N_{tb}$ phase of a homologue **CB-7-CB** observed in 1.6 micron cell at room temperature. The sample was quenched in liquid nitrogen and subsequently heated to room temperature. Slow recrystallization of the material is visible. In the inset the texture of the same sample after removing the cover glass showing system of densely packed focal conic defects.

The stripe periodicity is nearly equal the cell thickness, being 1.6, 3.1 and 5.3 micron in 1.6, 3.0 and 5.1 micron thick cell, respectively. The stripes are caused by the periodic modulation of local optical axis direction coupled to weak modulations of optical retardation (Fig. 2). The lowest retardation (corresponding to $\Delta n = 0.10$) is detected in regions in which director is along the rubbing direction, the regions in which the director is inclined from the rubbing have a higher optical retardation (corresponding to $\Delta n = 0.13$), being slightly lower than the retardation in the uniform N phase, suggesting some internal short wavelength structure, partially averaging the positions of the molecules. Apparently, in the regions of smaller retardation the optical axis is inclined from the surface plane. The tilting of the cell in respect to the light beam, along the strips, does not differentiate the retardation in these regions, neither influence the pattern obtained by diffraction of laser beam on the refractive index grating formed in the cell. Thus we can assume that in the neighbouring stripes of lower retardation the optical axis has the same direction of inclination from the surface.

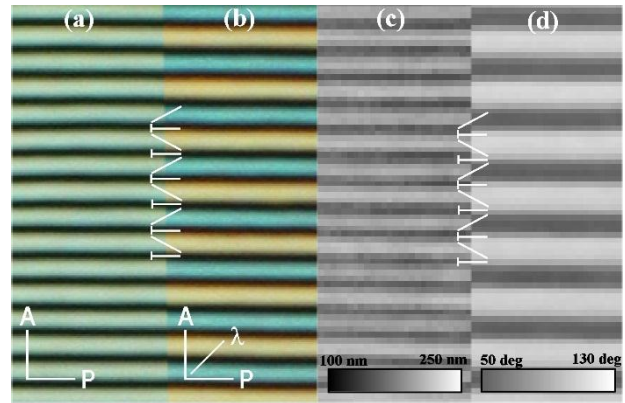

**Figure 2.** Optical stripe texture of N$_{tb}$ phase found for **CB-9-CB** in a 1.6 μm thick cell, uniformly rubbed in horizontal direction: (a) observed between crossed polarizers, (b) between crossed polarizers with a lambda plate inserted. Different colors indicate that the optical axis azimuthal angle alternates between stripes. Optical retardation (c) and azimuthal angle of optical axis (d) measured with Abrio imaging system. The retardation of optical axis changes periodically between 210 nm and 160nm and the azimuthal angle between 60 and 120 degree with respect to horizontal direction. Extinction regions in (a) correspond to minimal retardation and azimuthal angle 90 deg.

Such model of periodic optical axis modulations is consistent with additional observations: rotation of the cell between crossed polarizers shrinks the distance between every second neighbouring stripes. Upon lowering the temperature the inclination angle of optical axis modulations increases up to ~40 degrees, and as a result the amplitude of retardation modulations becomes more pronounced. The stripe texture is clearly stabilized by interactions with sample surfaces; upon removing the glass substrate, the stripes relax to arrays of pseudo-focal conic (FC) defects (Fig.1), the width of FC domains is about twice of the optical stripe periodicity. Presence of FC domains unambiguously proves the existence of an internal periodic structure of the nematic phase as such defects are characteristic for soft materials in which layering occurs, layers can be due to the positional ordering of molecules (smectics), short helices (chiral nematic), but also any other (e.g. splay-bend) periodic structures. The characteristic texture, with arrays of FC-like defects, enabled us to unambiguously distinguish the N$_{tb}$ phase in AFM measurements. Careful optical examination of samples frozen in liquid nitrogen and brought to room temperature usually showed large areas covered by the $N_{tb}$ phase, but also some nucleation centres of the crystalline phase were visible. It takes typically several minutes for homologues with n= 9 and 11 and less than one hour for n=7, to see coverage of the whole (0.25cm$^2$) sample area by solid crystal, the formation of crystalline phase was proved by XRD studies. Thus most of the AFM studies were made with the material **CB-7-CB**, for which the time interval, in which modulated nematic phase can be observed, is the longest. In the $N_{tb}$ phase a smooth surface over large areas (~100 x 100 μm$^2$) covered by focal conic defects (Fig. 3a) was observed.

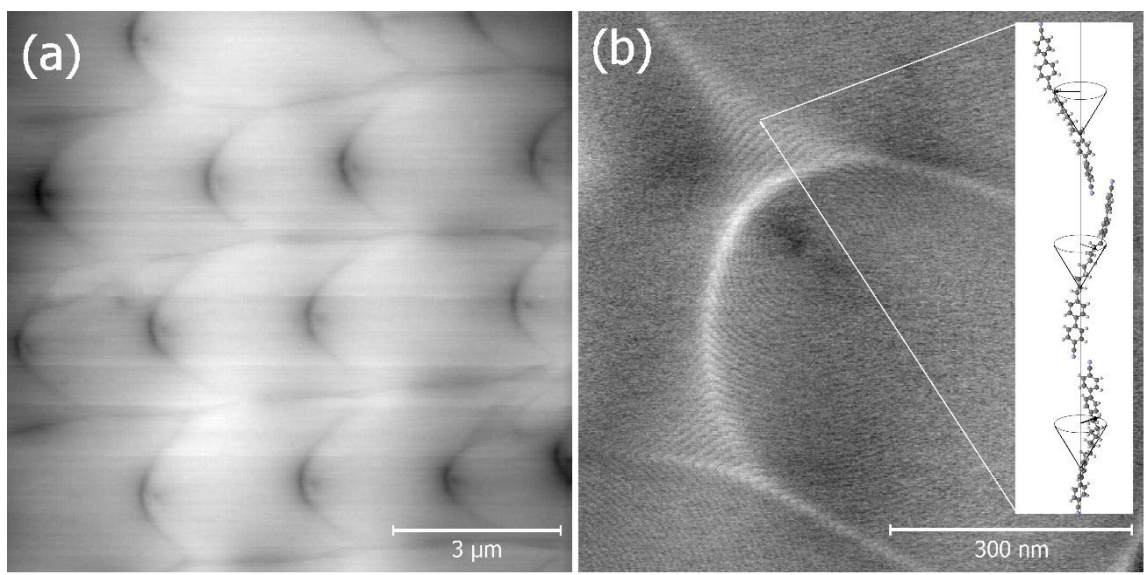

**Figure 3.** AFM images of the surface morphology of **CB-7-CB** sample in the Ntb phase with (a) arrays of focal conic defects and (b) small focal conic defect covered with a system of lines with ~8 nm periodicity. In the inset the molecular arrangement with three fold rotational symmetry present in crystallographic structure [ref. 17, reconstructed form CSD database, ref.cod. AZUGUX], that is most probably related to nanoscale periodicity of FCDs.

The shape of the defect lines in most places is parabolic, however elliptical defect lines were also observed. The weak depression of the surface around focal point and small inflation of surface around the parabolic (elliptical) defect line was detected (Fig. 3b).

Most of FC defects are oriented with their parabolic (elliptical) line nearly parallel to the sample surface, 'layers' are nearly perpendicular to the surface and the focal point is located close to the defect line, in 1.6 micron cell the distance is ~ 300 nm. The eccentricity parameter for elliptical domains ($e^2 = 1 - \frac{a^2}{b^2}$, where $a$ and $b$ are short and long semiaxes of the ellipse, respectively), deduced from the shape of FC domains and position of hyperbolic point, is large, $e^2 \sim 0.8$, that might suggest a large splay energy constant for the space modulated nematic [16]. Some of the focal conics are covered by a system of equidistant lines, having ~8 nm periodicity, corresponding to height differences of less than 0.5 nm (Fig. 3b). Under application of an a.c. electric field (~50$V_{pp}$/μm) perpendicular to the sample surface, the optical stripe texture was irreversibly converted to an optically non-birefringent texture (homeotropic alignment). Such samples after freezing were also examined with AFM, revealing that non-birefringent areas are made of densely packed toric domains (Fig. 4).

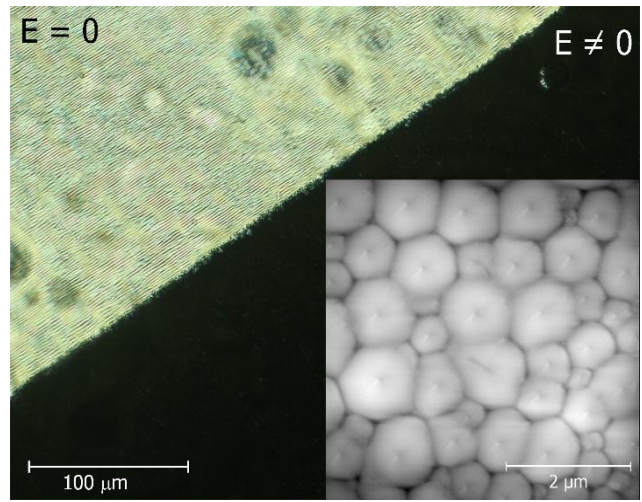

**Figure 4.** Change of optical texture of **CB-7-CB** in a 1.6 micron thick planar cell induced by application of an electric field perpendicular to the sample surface. Outside electrode area (E = 0) the stripe texture remained unaffected. The inset shows the morphology of the homeotropic part of the sample imaged by AFM, with a system with densely packed toric focal conic domains.

Apparently, under a strong electric field, reorientation of 'layers' takes place, FCDs are embedded into the system with layers mostly parallel to the surface (eccentricity of elliptical defect is close to 0). The areas covered by the solid crystal have very different morphology, in these regions nano-sized periodic structures were clearly visible (Fig. 5). Depending on the orientation of crystal planes toward the sample surface, different AFM images were registered, in some areas strongly curved layers are clearly detected, in others uniformly oriented crystallographic planes with a ~8nm periodicity were found. Knowing that the bi-phenyl **CB-9-CB** dimer crystalizes in a trigonal lattice with *P*3$_1$21 point symmetry [17], the layers could be identified as (001) crystallographic planes. It is quite striking that the crystal periodicity is very close to periodicities observed previously by TEM techniques in replica of freeze fractured samples [3,4] and to periodicity of tiny strips covering focal conic defects visible in AFM images (Fig. 3b).

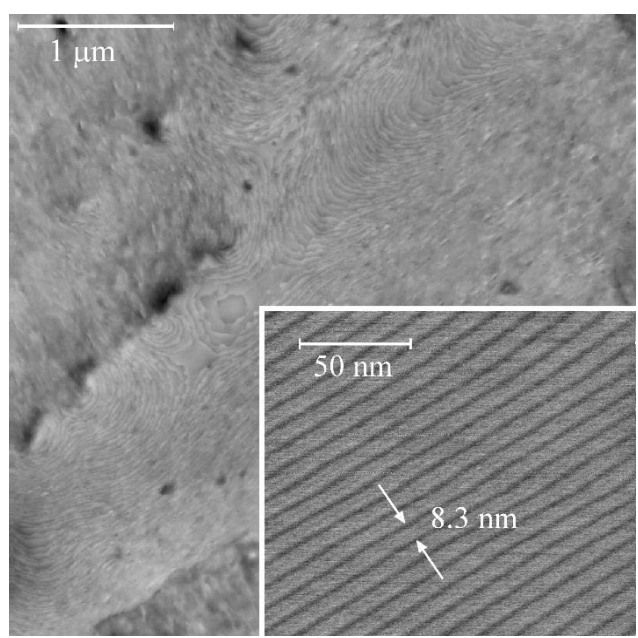

**Figure 5.** AFM image of the surface morphology of **CB-7-CB** in the crystalline phase with the clearly visible structure of submicron periodicity. The inset shows a high resolution image of the region with uniformly aligned crystal planes having a 8.3nm periodicity.

Crystalline and liquid crystalline areas of the samples were also distinguished by registration of force-versus-distance curves (Fig. 6), providing information on local material properties, such as elasticity, hardness, and adhesion, that are expected to be very different for a solid crystal and a liquid crystal [18].

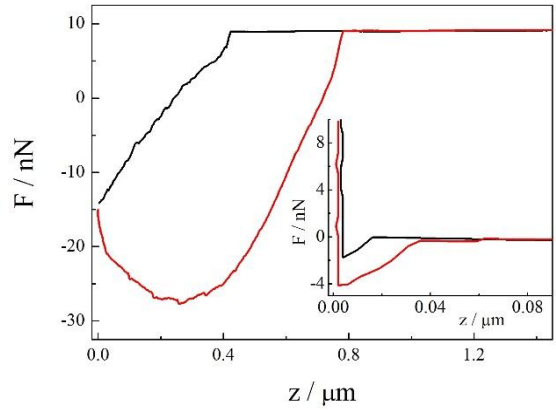

**Figure 6.** Force-distance curves for the $N_{tb}$ phase and (in the inset) crystalline phase regions formed in the sample of **CB-7-CB**; black curves are for approaching the surface, red for retracting AFM tip from the surface.

The force-versus-distance curve for crystalline areas shows the typical shapes for solid state, with well-defined contact point (at the distance ~20 nm) and linear slope below the contact distance. When the tip was retracted from the surface, the force decreased very rapidly due to inelastic deformation of the surface, and its weak adhesion. On the other hand liquid crystalline regions have very different characteristic, the contact point is less defined (~400 nm), and the large hysteresis between approaching and retracting scans is found, due to the strong surface adhesion in the fluid state.

In summary, our results show that ~8 nm regular periodic structures, observed previously by TEM imaging, and being ascribed to the helical twist-bend structure, in many cases might evidence the formation of a solid crystal. The investigated materials, **CB-n-CB**, have a strong tendency to re-crystallize ($N_{tb}$ is a monotropic phase), hence the $N_{tb}$ phase can only be observed for a short time interval at room temperature, even if the sample has been frozen in liquid nitrogen prior to the observation. Nevertheless, the formation of focal conic defects evidences unambiguously the presence of a short-wavelength spatially modulated structure of the lower temperature nematic phase. Moreover, even under a strong electric field the 'layered' structure of nematic phase is not destroyed, the electric field only reorients the 'layers' by moving the hyperbolic point to the centre of ellipse in focal conic domains, i.e., changing the eccentricity of FCDs. Whether the 8nm stripes, visible by AFM studies, are indicative of helicoidal structure is more ambiguous. The AFM (as well as TEM) method does not give the hint what is the nature of internal submicron scale structure. If indeed nematic phase is made of oblique helices, as suggested recently [3,4], the structure must be similar to that found for the crystalline phase of **CB-9-CB** [17], but lacking positional order i.e. with three fold symmetry and a small cone angle of ~ 20 deg. Such a structure would be reminiscent of the SmC$\alpha$ phase [19], but with short range positional order. Formation of short-wavelength helices averages molecular positions on nano-scale length and leads to optically uniaxial structure. In thin cells the spatial modulation of optical axis leads to appearance of stripe pattern, with periodicity driven by cell thickness. However, also the alternative explanation, that nematic phase has submicron, but larger than 8 nm periodicity has to be considered, while 8-nm structure visible in AFM studies reflects the surface freezing; the focal conics might be covered by a thin layer of either crystalline [20] or smectic phase [21].

*Experimental:* Optical examination of the characteristic textures of studied phases has been performed with a Zeiss Axio Imager A2m polarizing microscope equipped with a Linkam LTS-350 heating stage. For quantitative determination of sample birefringence and optical axis direction the CRI Abrio Imaging System integrated with microscope was used. Samples were prepared in glass cells with various thickness, 1.6 – 10 micron, having surfactant layers for either planar or homeotropic alignment. The same cells were also used for preparation of the samples for AFM studies – after the formation of a desired texture, the cell was immersed in liquid nitrogen, then brought to room temperature and broken. AFM images have been taken with Bruker Dimension Icon microscope, working in tapping mode at liquid crystalline-air surface. Cantilevers with a low spring constant, $k = 0.4$ $Nm^{-1}$ were used, the resonant frequency was in a range of 70-80 kHz, typical scan frequency was 0.1 Hz. The AF microscope was equipped with a camera, this allowed to monitor the investigated areas optically, so the crystalline and nematic regions in the samples can be distinguished unambiguously. XRD experiments were performed with Bruker GADDS system.

The **CB-n-CB** compounds were prepared by Hull group and re-synthesized by Warsaw group.

**Acknowledgements**

This work was financed by Foundation for Polish Science under program MASTER 3/2013. Z. A., C. W. and G.H.M. thank the EPSRC (UK) grant EP/J004480/1 and the EU for project funding through the projects EP/G030006/1 and 216025.

**References**
1. G. Ungar, V. Percec, M. Zuber, *Macromolecules* 1992, **25**, 75.
2. V. P. Panov, M. Nagaraj, J. K. Vij, Yu. P. Panarin, A. Kohlmeier, M. G. Tamba, R. A. Lewis, and G. H. Mehl *Phys. Rev. Lett.* 2010, **105**, 167801.

3. D. Chen, J. H. Porada, J. B. Hooperc, A. Klittnick, Y. Shen, M. R. Tuchband, E. Korblova, D. Bedrov, D. M. Walba, M. A. Glaser, J. E. Maclennan, and N. A. Clark *Proc. Natl. Acad. Sci.* 2013, **110**, 15931

4. V. Borshch, Y.-K. Kim, J. Xiang, M Gao, A Jákli, V. P. Panov, J. K. Vij, C. T. Imrie, M. G. Tamba, G. H. Mehl & O. D. Lavrentovich *Nat. Commun.* 2013, **4**, 2635.

5. A. Zep, S. Aya, K. Aihara, K. Ema, D. Pociecha, K. Madrak, P. Bernatowicz, H. Takezoe and E. Gorecka *J. Mater. Chem. C* 2013, **1**, 46.

6. C. S. P. Tripathi, P. Losada-Pérez, C. Glorieux, A. Kohlmeier, M-G. Tamba, G. H. Mehl and J Leys *Phys. Rev. E,* 2011, **84**, 041707.

7. V. P. Panov, R. Balachandran, M. Nagaraj, M. -G. Tamba, A. Kohlmeier, G. H. Mehl and J. K. Vij. *Appl. Phys. Lett.* 2011, **99**, 261903.

8. V. P. Panov, R. Balachandran, J. K Vij, M. G Tamba, A. Kohlmeier, G. H Mehl, *Appl. Phys. Lett.* 2012, **101**, 234106.

9. C. Meyer, G.R. Luckhurst, I. Dozov, *Phys. Rev. Lett.* 2013, **111**, 067801.

10. L. Beguin, J. W. Emsley, M. Lelli, A. Lesage, G. R. Luckhurst, B. A. Timimi, and H. Zimmermann *J. Phys. Chem. B,* 2012, **116**, 7940

11. J. W. Emsley, M. Lelli, A. Lesage, G. R Luckhurst, *J. Phys. Chem. B* 2013, **117**, 6547.

12. R. B. Meyer, Les Houches Summer School in Theoretical Physics, 1973. Molecular Fluids, ed. Balian, R., Weil, G., Gordon and Breach, New York, 1976, 271.

13. I. Dozov, *Liq.Cryst.* 2002, **29**, 483.

14. S. M. Shamid, S. Dhakal, J. V. Selinger, *Phys. Rev. E* 2013, **87**, 052503.

15. A. Hoffmann, A.G. Vanakaras, A. Kohlmeier, G.H. Mehl, D.J. Photinos arXiv: 2014, 1401.5445v2

16. M. Kleman, O.D. Lavrentovich, *Eur. Phys. J. E* 2000, **2**, 47

17. K Hori, M. Iimuro, A. Nakao, H. Toriumi, *J. Mol. Struct.* 2004, **699**, 23.

18. H.-J.Butt, B. Cappella, M. Kappl, *Surf. Sci. Rep.* 2005, **59**, 1

19. H. Takezoe, E. Gorecka, M. Cepic, *Rev. Mod. Phys.* 2010, **82**, 897

20. X. Z. Wu, E. B. Sirota, S. K Sinha, B. M. Ocko, M. Deutsch, *Phys. Rev. Lett.* 1993, **70**, 958

21. J. Als-Nielsen, F. Christensen, P.S. Pershan, *Phys. Rev. Lett.* 1982, **48**, 1107